\documentclass[conference]{IEEEtran}

\usepackage{cite}

\ifCLASSINFOpdf
   \usepackage[pdftex]{graphicx}
\else
   \usepackage[dvips]{graphicx}
\fi

\usepackage[cmex10]{amsmath}

\begin{document}
%
\title{Large Collection of Diverse Gene Set Search Queries Recapitulate Known Protein-Protein Interactions and Gene-Gene Functional Associations}

\author{\IEEEauthorblockN{Neil R. Clark$^{1,2,3}$, Avi Ma'ayan$^{1,2,3, *}$}
\IEEEauthorblockA{$^1$Department of Pharmacology and Systems Therapeutics, Icahn School of Medicine at Mount Sinai, \\
One Gustave L Levy Place, Box 1215, New York, NY, USA \\
$^2$Big Data to Knowledge (BD2K) Library of Integrated Network-based Cellular Signatures (LINCS) \\ 
Data Coordination and Integration Center (DCIC)\\ 
$^3$Mount Sinai Knowledge Management Center (KMC)
for Illuminating the Druggable Genome (IDG)\\
$^*$Corresponding author:
avi.maayan@mssm.edu
}

}

\maketitle

\begin{abstract}
Popular online enrichment analysis tools from the field of molecular systems biology provide users with the ability to submit their experimental results as gene sets for individual analysis. Such queries are kept private, and have never before been considered as a resource for integrative analysis. By harnessing gene set query submissions from thousands of users, we aim to discover biological knowledge beyond the scope of an individual study. In this work, we investigated a large collection of gene sets submitted to the tool Enrichr by thousands of users. Based on co-occurrence, we constructed a global gene-gene association network. We interpret this inferred network as providing a summary of the structure present in this crowdsourced gene set library, and show that this network recapitulates known protein-protein interactions and functional associations between genes. This finding implies that this network also offers predictive value. Furthermore, we visualize this gene-gene association network using a new edge-pruning algorithm that retains both the local and global structures of large-scale networks. Our ability to make predictions for currently unknown gene associations, that may not be captured by individual researchers and data sources, is a demonstration of the potential of harnessing collective knowledge from users of popular tools in the field of molecular systems biology.

Keywords - gene sets, big data, item set mining
\end{abstract}

\IEEEpeerreviewmaketitle


\section{Introduction}
Systems approaches to genome-wide molecular data are increasingly using gene sets, as opposed to individual genes, as the basic units of analysis. For example, in the field of molecular diagnostics, biomarker sets, as opposed to single gene biomarkers, are increasingly being applied \cite{wang2008gene}. Beyond molecular diagnostics, genomics, transcriptomics and proteomics experiments often identify gene sets that are differentially expressed, or gene sets that contain genetic variations associated with a phenotype. Using methods such as enrichment analyses \cite{chen2013enrichr,huang2009bioinformatics, subramanian2005gene}, common biological functions can illuminate prior knowledge originating from the collection of experimentally identified gene sets.

One approach to enrichment analysis is to take a set of experimentally identified genes and analyze these genes by comparing them to a library of gene sets with a known associated biological theme, e.g., members of cell signaling pathways, or targets of transcription factors.  Enrichr \cite{chen2013enrichr} is a popular online tool we developed to enable users to submit gene sets for this type of analysis. Since its launch in 2013, over 900,000 gene sets have been submitted to Enrichr by over 25,000 unique users. These gene sets originate from diverse experimental platforms, profiling genes and proteins at various regulatory levels of mammalian molecular data.

We hypothesize that the gene set submitted to Enrichr may contain patterns and structure that could reveal novel associations between genes, gene modules and molecular biological mechanisms. The large size of this data set and the diversity of the scientific community from which it originates potentially provide a unique perspective that may reveal new insights.

In combinatorial mathematical terms, the nearly 1 million gene sets that were submitted to Enrichr comprise a family of sets, i.e., a collection of subsets from the set of all genes in the human, mouse and rat genomes. A common approach to the analysis of this form of data is to treat each set as an item-set in analogy to a market basket, and identify frequently occurring combinations of items \cite{brin1997dynamic}. An alternative has been proposed in which item-sets that are logically related are identified \cite{kumar2012logical}. The identification of logical item-sets is particularly relevant for the gene sets submitted to Enrichr because this approach takes into account rare items, and because biological knowledge is incomplete, logically associated groups of items may only be partially represented in any given basket. These are important properties, as we aim to identify functional modules even for rarely occurring genes. We expect that due to partial information contained within an individual query, not all members of a relevant functional module will be present in any given gene set submitted for analysis with Enrichr. This partial information is expected to produce high level of false positive associations. Due to the expected high false positive rates, a degree of noise-filtering may also improve the accuracy of  such analysis.

Here we provide an interpretable picture of the global properties of a large collection of the lists submitted to Enrichr by thousands of users. To accomplish this we employ the method of logical item-set to extract gene modules. As a consequence, we infer a network of relationships between genes and visualize this network using a novel edge-pruning method we devised. Next, we examine the degree of compatibility of our findings with annotated gene set libraries such as the Human Gene Atlas or the Gene Ontology (GO), as well as with known protein-protein interactions.

\section{Results}
By the middle of 2015, the Enrichr gene set queries consisted of 172,798 gene sets submitted from 5,114 unique internet protocol (IP) addresses. Our analysis began by applying a number of filters. First, we removed any query that did not correspond to an official gene symbol. Next, we ensured that no individual users dominated in their contributions to the data set by removing all entries from users that contributed many gene sets (see methods for more details). After this strict cleaning criteria, the result was a family of 19,196 gene sets, composed of 27,770 genes supplied from 3,308 unique IP addresses of unique users of Enrichr.

\subsection{Enrichr gene set library metrics}
\begin{figure}[t]
\centering
\includegraphics[width=3.5in]{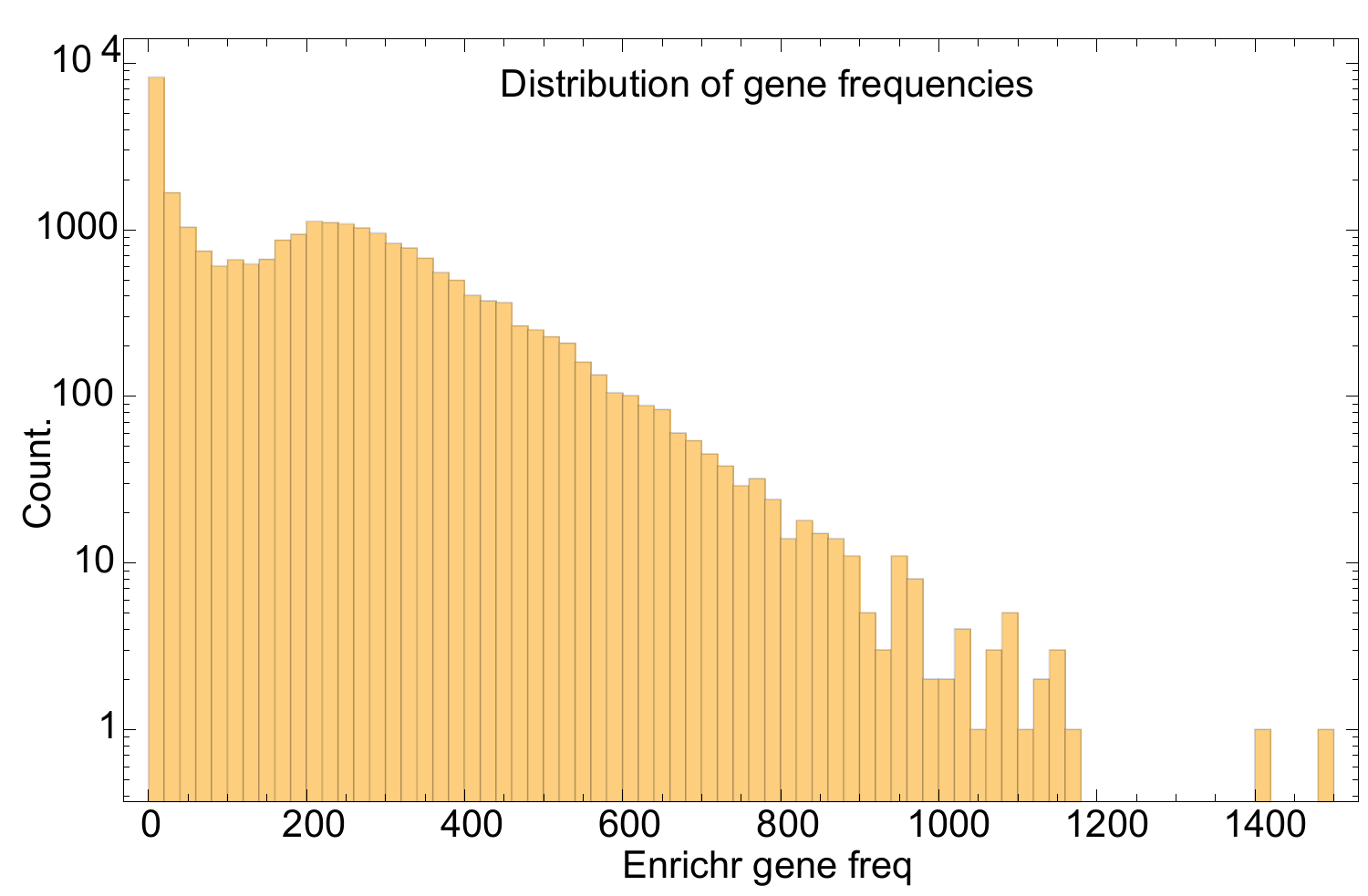}
\caption{ The distribution of the occurrences of unique gene symbols in Enrichr gene set queries.}
\label{fig-genefreqdist}
\end{figure}

The distribution of gene frequencies is surprisingly complex (Fig. \ref{fig-genefreqdist}). The distribution approximately decays exponentially from frequencies greater than ~300; however, the lower frequencies are characterized by a bimodal distribution. The large mode at the lowest frequencies has an overrepresentation of uncharacterized genes, gene names with no functional associations or known protein-protein interactions; but even discounting those genes, this peak remains relatively large. The lower frequencies are also characterized by a mode at around 200 occurrences. The frequency of occurrence of genes in submissions to Enrichr has a characteristic frequency that decays exponentially, as opposed to being characterized by a power law. Given the predominance of scale-free distributions, it is perhaps surprising that there is a definite scale for frequencies of gene occurrence in this data set. 

\begin{figure}[t]
\centering
\includegraphics[width=3.5in]{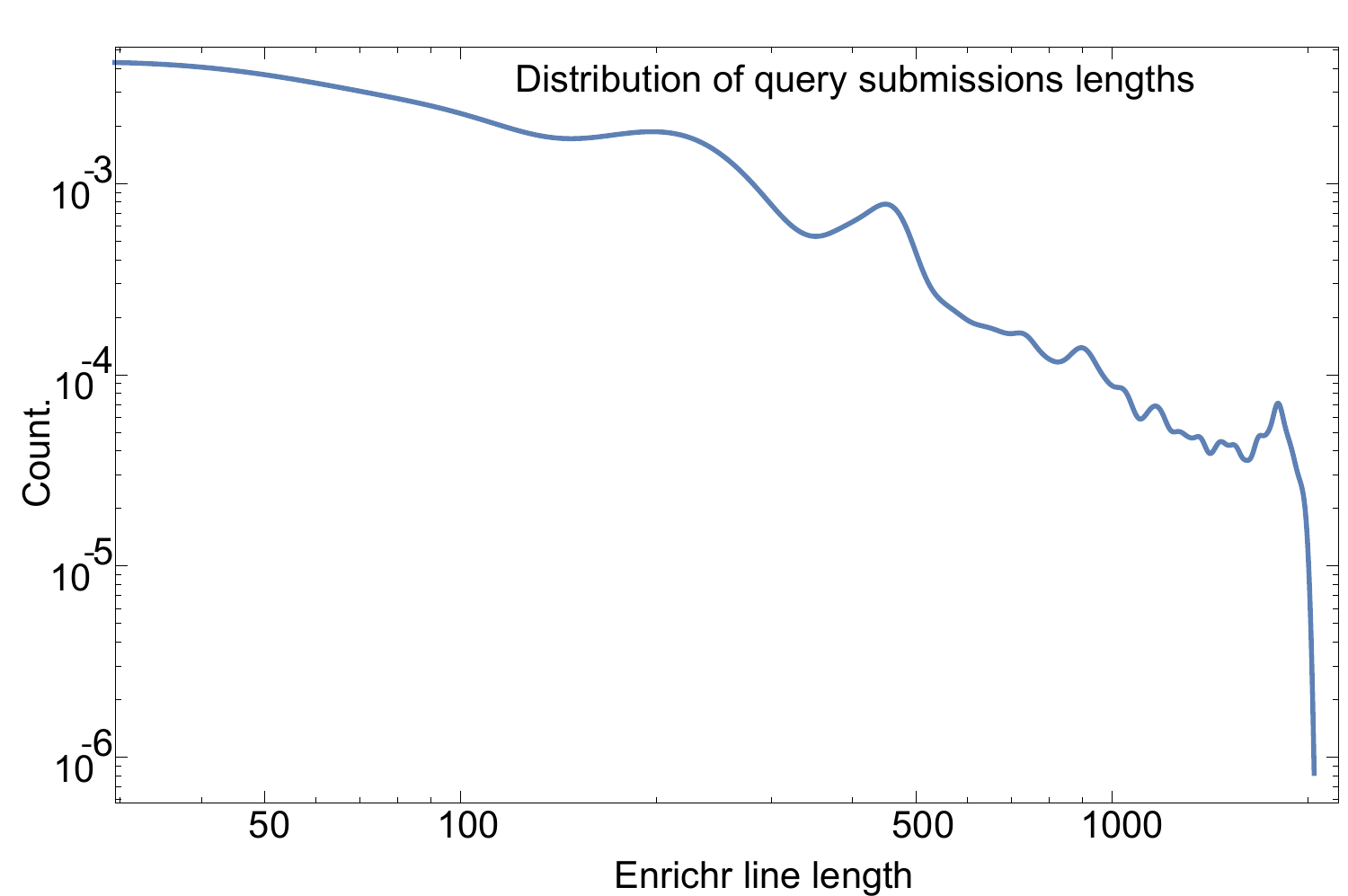}
\caption{ The distribution of the size of each gene set query submitted to Enrichr.}
\label{fig-linelengths}
\end{figure}

The distribution of the lengths of gene sets submitted to Enrichr approximately fits a power-law, with a notable peak at about 200 and 400 genes (Fig. \ref{fig-linelengths}). The distribution of the number of submissions to Enrichr per user IP address also follows a power-law, suggesting that most users submit only a few lists, but there is a substantial body of heavy users.

\subsection{Gene-gene and gene-set/gene-set network inference}
The Enrichr collection of gene set queries can be transposed, such that for each gene there is an associated set of Enrichr submission queries. To explore the structure in this data, and also to potentially form the basis for an informative decomposition of the data, we can infer networks of similarities between gene pairs and, in addition, networks of Enricr queries. We employ the methods developed in \cite{kumar2012logical} to infer noise-filtered networks of associations between these entities.

\begin{figure*}[!t]
\centering
\includegraphics[width=6in]{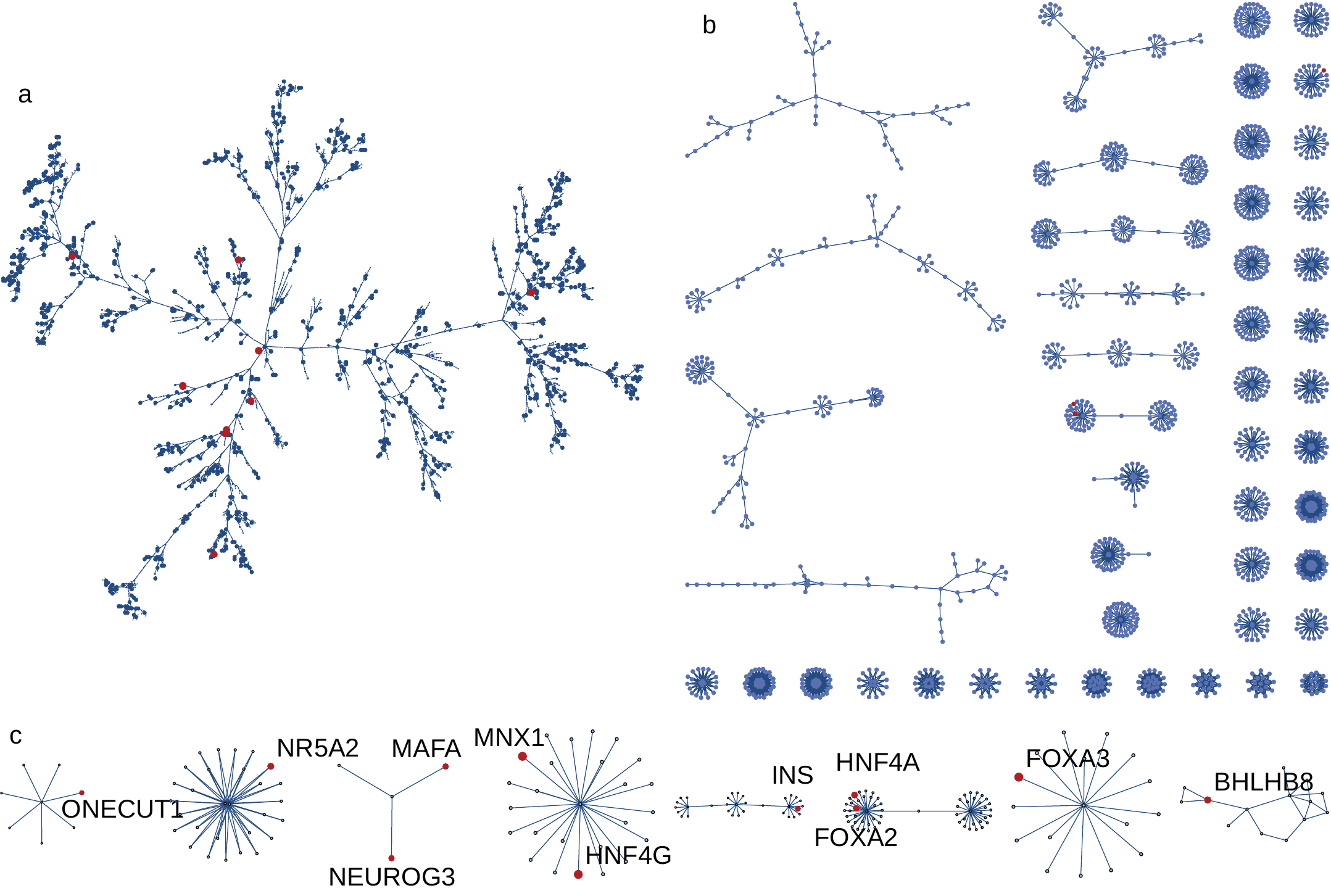}
\caption{The edge-pruned network of gene associations inferred from Enrichr queries. (a) The global view of the network. (b) A selection of the largest local network structures. (c) As an example of one possible use of this network, we highlight genes known to be associated with adult-onset diabetes (KEGG) and illustrate the local network structure that includes at least one of these genes.}
\label{fig-genenetwork}
\end{figure*}

In order to visualize the network of gene associations, we employ an edge-pruning algorithm that results in an interpretable representation of the global network structure while preserving local features of the network topology (Fig. \ref{fig-genenetwork}a). A selection of the largest local structures, preserved by the edge pruning algorithm, is shown in Fig. \ref{fig-genenetwork}b. One potential use of this network is to make predictions of novel associations based on prior knowledge. As an example of this, we highlight genes known to be associated with Adult Onset Diabetes (KEGG) (Fig. \ref{fig-genenetwork}c). Illustrating local network structures that include at least one of these genes suggests local structures as candidate novel predictions for genes likely associated with the adult-onset diabetes pathway.

One of the local structures contains the adult-onset diabetes Genes MAFA and NEUROG3 (KEGG) along with two other genes, UTS2R and FLJ45717, that are not identified in KEGG as being associated with diabetes. However, UTS2 and its receptor UTS2R have been reported \cite{jiang2008comparative} to be involved in glucose metabolism and insulin resistance, which lead to the development of type-2 diabetes in humans. There is no known association of FLJ45717 with diabetes; however, as this gene forms the center of a star graph local structure, we consider it a novel candidate for involvement in adult-onset diabetes.

\begin{figure*}[!t]
\centering
\includegraphics[width=6in]{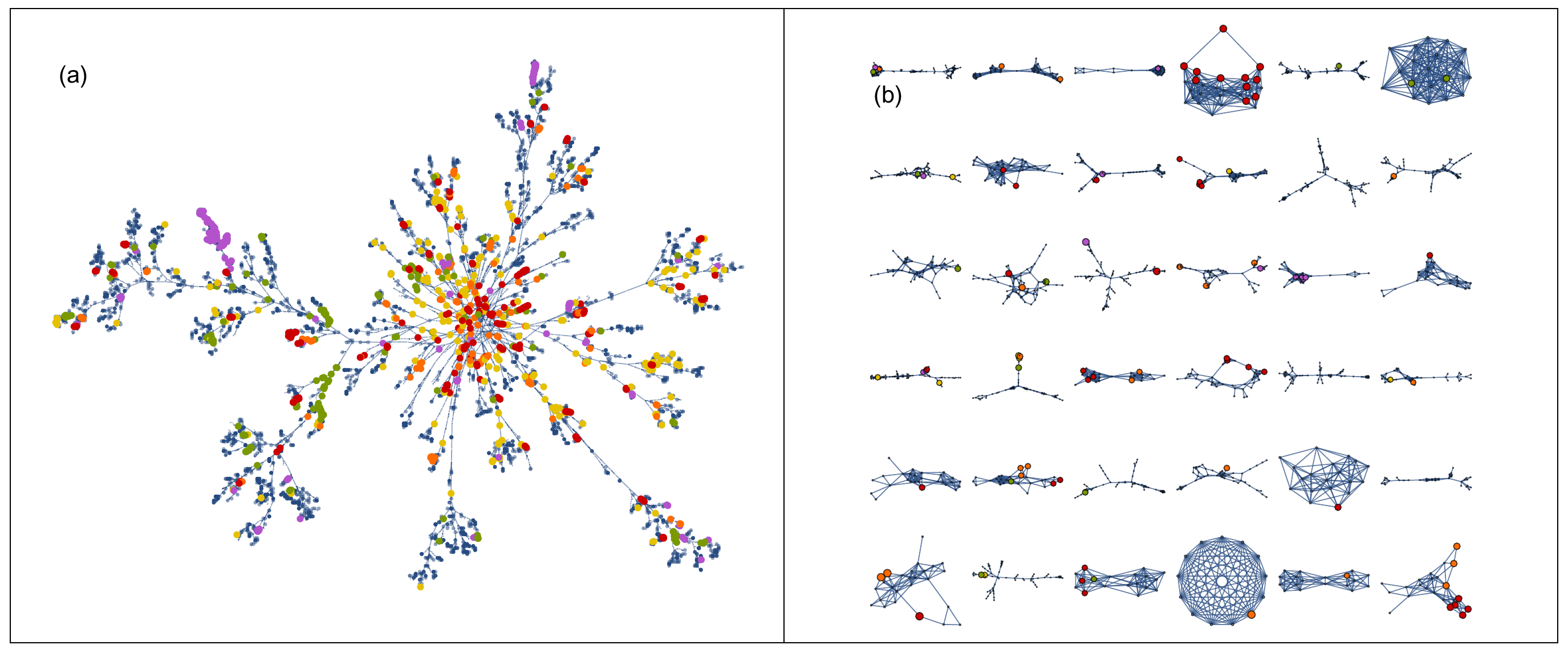}
\caption{The edge-pruned network of associations between query submissions to Enrichr. (a) The global view of the network. (b) A selection of the largest local structures. (c) The gene sets submitted by the five Enrichr users with the largest number of submissions are highlighted, each with a different color.}
\label{fig-linenetwork}
\end{figure*}

We can also apply the same approach to the network of associations between gene set queries submitted to Enrichr (Fig. \ref{fig-linenetwork}). This network appears to show a degree of clustering by user; this perhaps reflects individual users special research interests, or a common method of data acquisition.

\subsection{Comparing the clustering of Enrichr networks to random scale-free networks}
The clustering coefficients of the gene and gene-set networks are 0.43 and 0.17, respectively. In order to gauge the significance of this, we compared these clustering coefficients to the typical clustering coefficients for a random scale-free network generated by the Barabasi-Albert random scale-free graph model. Artificial shuffled networks of a similar size and degree distribution generated by the Barabasi-Albert model have a typical clustering coefficient of 0.0026 and 0.0054 respectively. Hence, we see that the Enrichr queries gene, and gene-set, networks have a clustering coefficient which is orders of magnitude greater than would be expected if the networks were randomly scale-free by the Barabasi-Albert model.

\subsection{Examining the gene network for recovery of known protein-protein interactions}
Next, we asked whether the gene-gene association network created from the Enrichr queries can be used to predict physical protein interactions. We used the PSICQUIC database of protein-protein interactions (PPIs) and examined the statistical significance of the overlap between the PSICQUIC PPI network and the gene-gene network inferred from Enrichr submissions. After applying a threshold of 0.3 for the similarity of a normalized point-wise mutual information for the gene-gene association network, two binary matrices remained.

In the first test of the significance of the similarity between these two binary matrices, we count the number of non-zero entries in each matrix and the number of elements that are nonzero in both matrices. Based on a null distribution whereby the matrices are randomly permuted, we can use the hypergeometric distribution to quantify the significance of the overlap between the two matrices.

There are $n_1=13,586$ genes that are in both the Enrichr queries network and the PSICQUIC PPI network. With a threshold on the normalized point-wise mutual information of 0.05, there are $n_1=99,758$ edges in the Enrichr network and $n_2=156,388$ known PPIs. There are $n_k=2,763$ edges that are present in both the Enrichr network and the PSICQUIC PPI network. In this regime, the Poisson approximation to the hypergeometric distribution is applicable. With a total number of possible edges given by $n_g (1-n_g)$, the rate parameter for the Poisson distribution is given by:
\begin{equation}
\frac{n_1 n_2}{n_g (1-n_g)}=169.1
\end{equation}
Under the null hypothesis that edges are randomly assigned, we expect a mean number of shared edges to be 169.1 with the standard deviation of approximately 13.0. Consequently, the observed value of $n_k=2,763$ is about 200 standard deviations greater than the expected value. Therefore, under the null hypothesis, the number of edges present in both the Enrichr gene network and the PSICQUIC PPI network is extremely statistically significant.

It is conceivable that this result is due to users submitting gene set queries containing proteins known to have many interactors, and hence the null distribution we used for the calculation would be inappropriate and the result invalid.

One way to account for this is to condition the test on the frequency of each gene. To do this, we perform a separate hypergeometric test for each gene, and correct for multiple hypothesis testing. There are 899 proteins that have at least one interaction that overlaps with the Enrichr network; when corrected for multiple hypotheses testing, with a False Discovery Rate of 2

Another possible explanation for the apparently significant overlap between edges in the Enrichr gene network and the PSICQUIC PPI network is that there are a significant number of user queries containing genes with known PPIs. To address this, we recovered the Enrichr lists that contribute to the prediction of at least one PPI and counted the number of proteins in each query list for which there is another protein in the query with which it interacts. If a user submitted a list of proteins with known protein interactions, then the ratio of the counted proteins to the length of the query will be large. We note that the actual ratio is less than 1\%, which suggests that users are not submitting gene sets with known PPIs to Enrichr to any significant degree. This suggests that the significant overlap in the edges in the gene network inferred from Enrichr submissions is predictive of known and potentially novel protein-protein interactions.

\begin{figure}[t]
\centering
\includegraphics[width=3.5in]{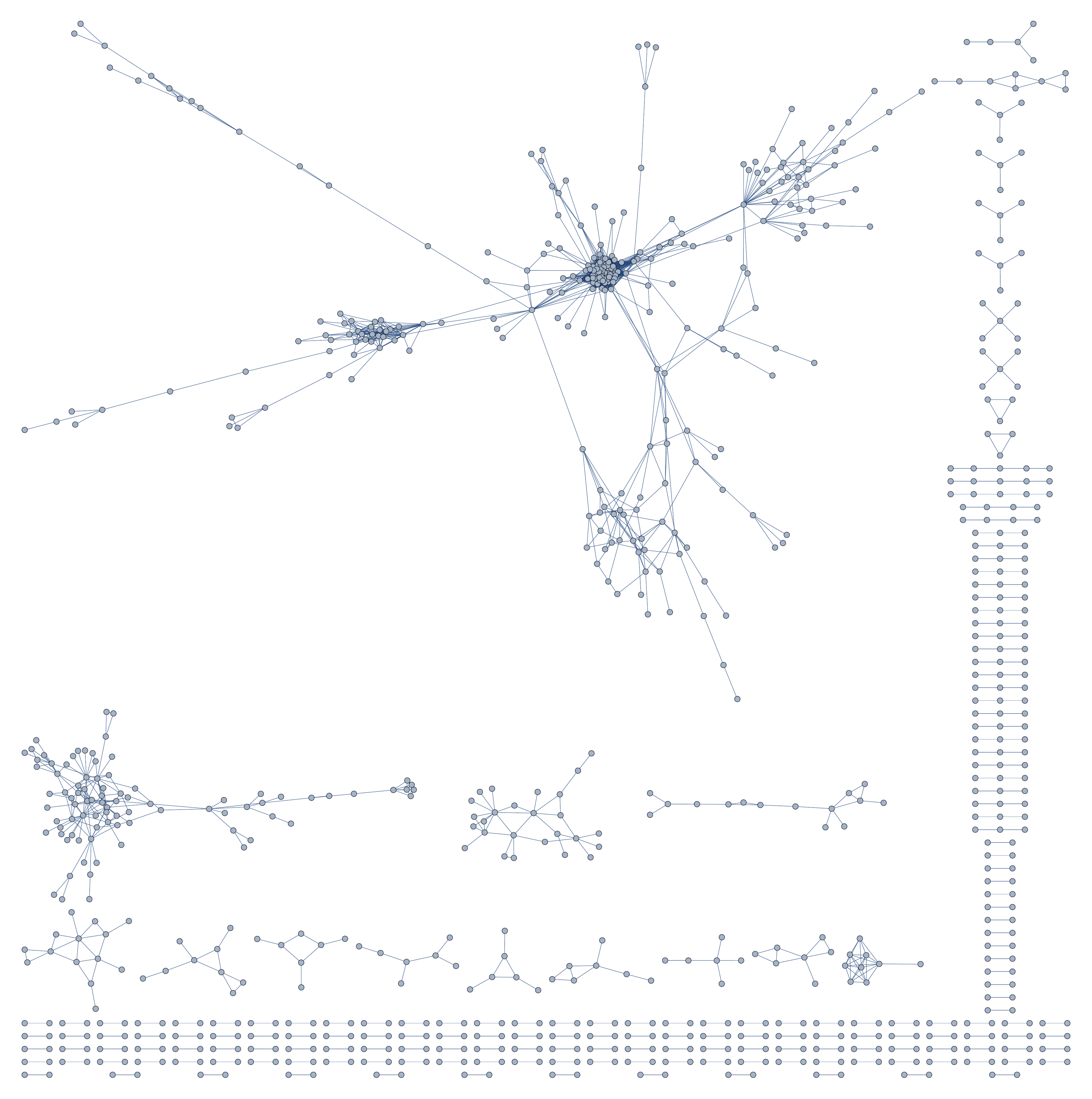}
\caption{The part of the network of the gene-gene associations inferred from Enrichr that is also supported by known protein-protein interactions from PSICQUIC.}
\label{fig-rePPI}
\end{figure}

We show the network of edges between genes that are present in both the Enrichr-inferred gene network and the PSICQUIC PPI network (Fig. \ref{fig-rePPI}).

\subsection{Evaluating the recapitulation of prior knowledge about gene functions with the gene-gene association network inferred from Enrichr queries}
We took a collection of 28 annotated gene set libraries that represent prior knowledge of associations between genes and their functions from a range of biological themes, including Gene Ontology, mouse phenotypes, protein complexes, histone modifications and DNA binding. We addressed the question of the extent to which the information in these gene set libraries is recapitulated by the gene-gene association networks inferred by Enrichr queries.

In order to test this for a single gene set, we mapped the genes in the set to nodes in the Enrichr queries gene-gene association network, and then calculated the mean point-wise mutual information between each pair of genes. In order to assess the significance of this collective measure of similarity, we numerically calculated a null distribution by randomly choosing a set of nodes in the Enrichr queries' network with the same cardinality as the gene set in question. The random choice was weighted by the overall frequency of the gene in the gene set library from which it originates. This is equivalent to generating a null distribution based on random permutation of the gene set library repeated many times. By comparing the actual collective similarity to the null distribution, we calculate a significance p value for the collective similarity of the gene set in the Enrichr queries' gene-gene association network. A small p value indicates that the members of the gene set in question are significantly more similar to each other in the network inferred by Enrichr queries than would be expected if the gene set library from which the set was originated was randomly permuted. We interpret a small p value as indicating that the network inffered by the Enrichr queries has recovered the gene set in question. The list of all gene set libraries examined (Table \ref{table-gmtdata}).

\begin{table}
\caption{Gene set libraries used to assess recovery of functional associations by the network inferred from the Enrichr queries.}
\label{table-gmtdata}
\centering
\begin{tabular}{l c c}
	Gene Set Library& \# of Sets &Mean Set Size\\
	\hline

BioCarta pathways	&	249	&	18 \\
Cancer Cell Line Encyclopedia	&	967	&	176\\
ChEA	&	240	&	1456\\
Chromosome location	&	386	&	85\\
CORUM	&	1673	&	5\\
Gene Ontology Biological Process	&	941	&	78\\
Gene Ontology Cellular Component	&	205	&	172\\
Gene Ontology Molecular Function	&	402	&	122\\
GeneSigDB	&	2139	&	127\\
Genome Browser PWMs	&	615	&	275\\
HMDB Metabolites	&	3906	&	47\\
Human Gene Atlas	&	84	&	450\\
KEA	&	474	&	37\\
KEGG pathways	&	200	&	48\\
MGI MP top 3	&	71	&	717\\
MGI MP top 4	&	476	&	202\\
microRNAs	&	222	&	155\\
Mouse Gene Atlas	&	96	&	660\\
NCI60	&	93	&	343\\
NURSA-IP-MS	&	1796	&	158\\
OMIM disease genes	&	90	&	25\\
OMIM Expanded	&	187	&	89\\
Pfam-InterPro-domains	&	311	&	35\\
PPI Hub Proteins	&	385	&	247\\
Reactome pathways	&	78	&	73\\
TF PPIs	&	290	&	79\\
VirusMINT	&	85	&	15\\
WikiPathways pathways	&	199	&	39\\
	
\end{tabular}
\end{table}

\begin{figure}[t]
\centering
\includegraphics[width=3.5in]{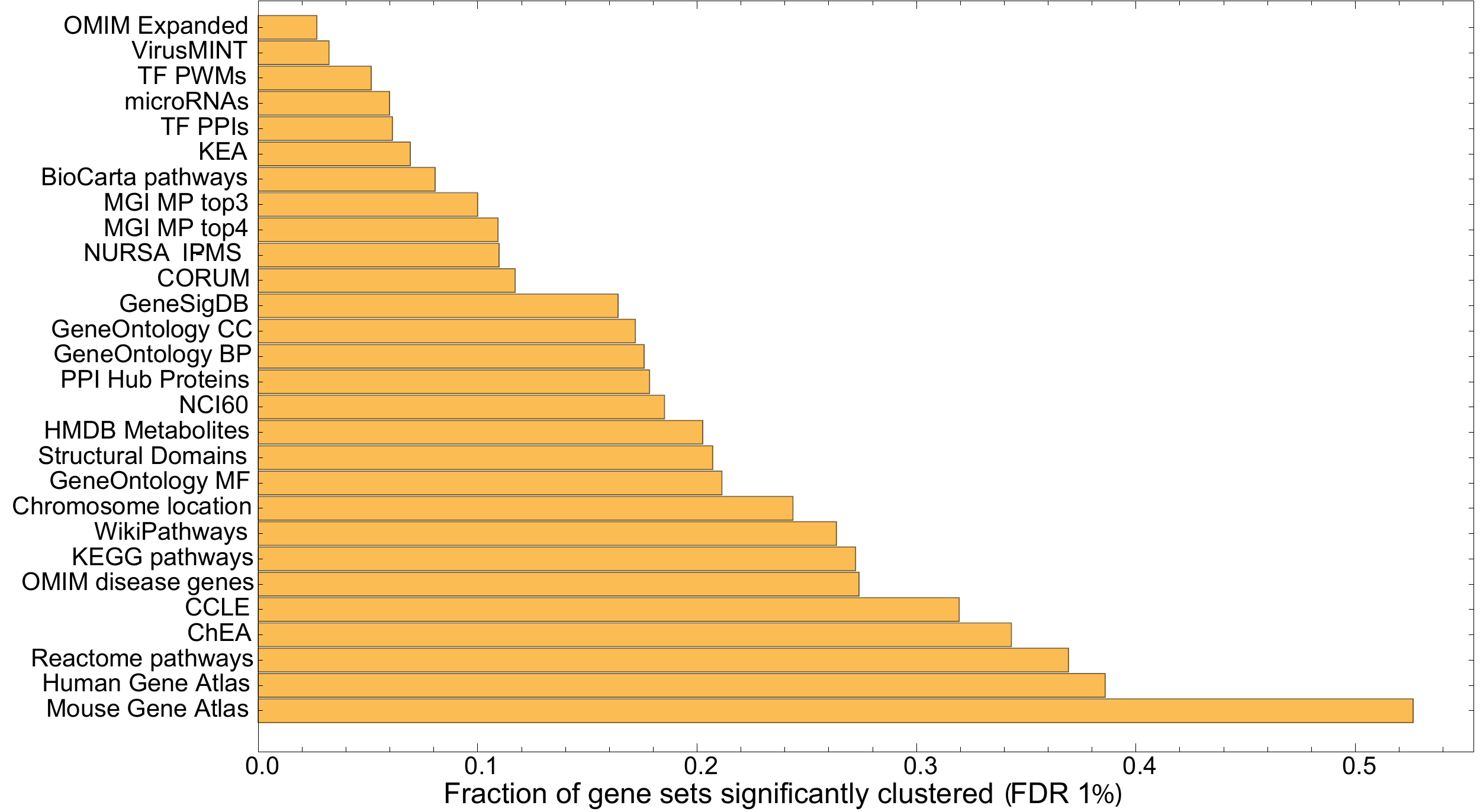}
\caption{The proportion of gene sets in each library that are significantly clustered in the gene-gene association network inferred from queries to Enrichr.}
\label{fig-reGMT}
\end{figure}

After correcting for multiple hypotheses testing and setting a false discovery rate threshold of 1\%, we counted the number of gene sets in each library that are significantly recovered in the Enrichr queries' gene-gene network (Fig. \ref{fig-reGMT}).

We observed that for some libraries, over 30\% of the gene sets in the library are recoverable by the Enrichr queries' gene-gene association network. 
The libraries containing data from genome-wide gene expression profiling have the most overlap, followed by ChEA, which has lists of genes associated with transcription factors from ChIP-seq profiling. Next are pathway libraries such as reactome, KEGG and WikiPathways. This rank order of libraries likely reflects the nature of most submitted lists to analysis with Enrichr. These query submissions are likely dominated by differentially expressed genes from genome-wide mRNA profiling. The recovery of annotated pathways suggests that the inferred gene-gene association network likely contains new and more complete pathway knowledge.      

\section{Discussion and Conclusions}

In this study, we analyzed for the first time, a large collection of gene set queries submitted by the many users of the popular enrichment analysis tool Enrichr. We examined the distribution of occurrence of genes, the length of submitted gene sets, and the distribution of submissions by individual users. We show that the distribution of gene occurrence in queries is complex, with some genes appearing often in submitted lists. The length of gene set queries peaks at 200-400 genes per query, and the distribution of submitters follows a power law. By constructing a gene-gene association network of co-occurrence, we were able to show that such a network captures known protein-protein interactions and functional associations much more frequently than by chance. This means that the collective data from thousands of queries potentially holds new knowledge about the local and global structure of the human functional interactome. As more submissions accumulate, it is expected that such predictions will be refined and improved. While experimental validation of these predictions is outside the scope of this present work, the gene-gene association network resource we developed here can be used to prioritize predictions for general and specific hypotheses that can be tested experimentally and systematically drive rational research explorations. The reuse of privately submitted gene sets by the Enrichr users should be handled with caution. The analysis that we conducted here and the gene-gene association networks we constructed keep the identities of users private. Furthermore, we do not provide the gene sets used for our analysis publicly.   

\section{Methods}
\subsection{Data Preprocessing}
As of June 25th 2015, 172,798 queries of gene sets were submitted to Enrichr from 5,114 unique IP addresses. In order to analyze the global structure of these gene sets, preprocessing was necessary because the raw data include multiple instances of the example data set supplied by the Enrichr website for demonstration purposes, special characters and strings that cannot be interpreted as gene lists caused by erroneous input from users, and large collections of queries from individual users who utilized the Enrichr API. These entries were removed so that the resulting dataset is representative of the entire community of Enrichr users and is not dominated by any individual user or small subset thereof.

The first filter was applied to retain only word strings that are members of a reference set of 39920 standard human, mouse and rat gene names. The resulting sets of genes corresponding to each user input list was then subjected to a sequence of subsequent filters.

Use of the Enrichr API enables programmatic access to Enrichr. This resulted in a minority of users submitting a large number of gene lists. In order for the data set to be balanced and representative of the entire community of Enrichr users, and not dominated by a small minority of users, we filtered out gene lists associated with IP addresses from which large numbers of lists were received.

The results of 11,579 differential expression analyses were received from the GEO2Enrichr Chrome Extension, which submits three gene sets for every differential expression computation for analysis with Enrichr: upregulated, down regulated and combined gene sets, respectively. Only the combined gene sets were retained in this analysis.

Gene sets that contained more than 2,000 genes were rejected on the basis that they were non-specific and incurred undue computational expense in subsequent analysis. 

Finally, all instances of the demonstration gene set from the Enrichr website were removed. The result was 19,196 gene sets composed of 27,770 genes supplied from 3,308 unique IP addresses.

\subsection{Network Inference}
We employ the methods described in \cite{kumar2012logical} to reconstruct networks from the filtered gene sets. 

The most basic approach to construct a network from a large collection of gene sets is to define a distance matrix between the gene sets and examine the resulting distance matrix. The most obvious similarity measure between two gene sets A and B is the Jaccard index:
\begin{equation}
J(A,B)=\frac{|A\cap B|}{|A \cup B|}
\end{equation}
Let the set of all elements be V, and the family of sets be M, then:
\begin{equation}
M=\left\{m^{(n)}=\left\{m_l^{(n)}\right\}_{l=1}^{L_n }\subset V \right\}
\end{equation}
and the co-occurrence counts of pairs of genes is given by:
\begin{equation}
\phi(\alpha,\beta)=\sum_{n=1}^N \delta(\alpha \in m^{(n)}) \delta(\beta \in m^{(n)})
\end{equation}

where $\delta(bool)$ is a Dirac delta function defined such that:
\begin{equation*}  
\delta(bool)=
	\begin{cases} 
		1 & \text{if} bool=True, \\
		0 & \text{Otherwise}    
	\end{cases}
\end{equation*}

The marginal counts are then defined in terms of the co-occurrence counts as:
\begin{equation}
\phi(\alpha)=\sum_{\beta \in G, \alpha \ne \beta } \phi (\alpha, \beta) 
\end{equation}

which is the number of all co-occurrences. Finally, the total number of co-occurrences is given by:
\begin{equation}
\phi_0=\frac{1}{2} \sum_{\alpha \in V} \phi(\alpha) =\frac{1}{2} \sum_{\alpha \in V}\sum_{\beta \in V} \phi(\alpha, \beta) 
\end{equation}
Then the co-occurrence probability is given by:
\begin{equation}
P(\alpha, \beta)=\frac{\phi(\alpha,\beta)}{\phi_0} 
\end{equation}
and the marginal probability is given by:
\begin{equation}
P(\alpha)=\phi(\alpha)/phi_0 
\end{equation}
The Jaccard index can now be defined in these terms as:
\begin{equation}
\psi_{jac}=\frac{P(\alpha, \beta)}{P(\alpha)+P(\beta)-P(\alpha,\beta))}
\end{equation}

When the collection of gene set queries is transposed such that the genes are the labels and the queries are the members of sets, the Jaccard index can be applied.

We can also use the following distance measures:
\begin{itemize}
	\item Cosine Distance: \newline $\psi_{cos}=P(\alpha,\beta)/\sqrt{P(\alpha)P(\beta) }$
	\item Point-Wise Mutual Information: \newline $\psi_{pmi}=\text{max}\left\{ 0,Log(P(\alpha,\beta)/P(\alpha)P(\beta) )\right\}$
	\item Normalized Point-Wise Mutual Information:\newline $\psi_{nmi}=-\psi_{pmi}/(Log(P(\alpha,\beta)))$
\end{itemize}
We apply a noise filter by iteratively applying a threshold for similarity: the pair counts for gene pairs that do not pass the threshold are set to zero, and the similarity matrix is recalculated.

Throughout the rest of this analysis, we only use the normalized point-wise mutual information as a measure of similarity between genes for the Enrichr queries data.

\subsection{Network visualization}
Because the resultant networks are densely connected, direct visualization of the graph is not interpretable. Edge pruning has been proposed for the simplification of graphs to aid in generating interpretable visualizations of their structure \cite{zhou2010network}. We find that such edge pruning algorithms, applied to very large graphs, are able to represent the global structure at the expense of the local structure. To potentially improve this, we employed an edge pruning approach that is able to capture both global and local structure in the graph.

In the initial step, a set of local structures is derived by varying the threshold for similarity and identifying connected graph components at the point at which they disconnect from the giant component. Such point is defined as any connected component that contains more than 10\% of all nodes or has an absolute number of nodes greater than 100 nodes; this number is chosen based on the visualizability properties of the resulting local structures.

The local structures represent local regions of the network that are distinguishable from the rest of the network at an appropriate scale of similarity. As such, they depict the similarity relationships between the nodes and their neighborhood in the network.

Once the local structures have been identified, we employ a naive edge-pruning algorithm. We do not prune edges that are members of the local structures, thereby preserving them in the final simplified network. The number of edges that are available to be pruned is equal to the difference between the total number of edges in the original network, $n_e$,  and the number of edges that are part of the local structures, $n_{e;ls}$. The pruning process preserves the connectivity of the network, so after the maximum degree of pruning, the minimum number of edges remaining must be equal to:
\begin{equation}
n_{ls}+n_{e;ls}
\end{equation}
Accordingly, the maximum number of edges that can be pruned while preserving the connectivity of the network is:
\begin{equation}
n_e-n_{e;ls}-n_{ls}
\end{equation}
The iterative process by which edges are pruned is as follows:
\begin{enumerate}
	\item Set a counter $i=0$
	\item Identify the edge with the lowest similarity measure
	\item If removing the edge does not cut the network, then remove it.
	\item $i\rightarrow i+1$
	\item If $i<\gamma(n_e-n_{e;ls}-n_{ls})$ then return to step 1; otherwise, stop.
\end{enumerate}

\section*{Acknowledgment}
We would like to thank Dr. Kathleen M. Jagodnik for providing useful comments and copyediting.
Funding: This work was supported in part by grants from the NIH: R01GM098316, U54HG008230 and U54CA189201 to AM.



\bibliographystyle{IEEEtran}

\bibliography{EnrichRSets-bibliography-01}

%




\end{document}